\documentclass[11pt,onecolumn,floatfix,superscriptaddress,nofootinbib]{revtex4-1}

\usepackage[english]{babel}
\usepackage{amsmath}
\usepackage{graphicx}                   
\usepackage{epsfig}
\usepackage{amsmath,amssymb}            
\usepackage{hyperref}                   
\usepackage{epstopdf}
\usepackage{placeins}
\usepackage{upgreek}
\usepackage{slashed}
\usepackage{color}
\usepackage{footmisc}
\usepackage{footnote}
\definecolor{orange}{rgb}{1,0.5,0}
\definecolor{brown}{rgb}{0.65, 0.16, 0.16}
\definecolor{phlox}{rgb}{0.87, 0.0, 1.0}

\graphicspath{{figs/}}                  
\usepackage{ulem,xcolor}

\def\frac#1#2{{\textstyle{{#1}\over {#2}}}}

\def\lsim{\mathrel{\rlap{\lower4pt\hbox{\hskip1pt$\sim$}}
    \raise1pt\hbox{$<$}}}
\def\gsim{\mathrel{\rlap{\lower4pt\hbox{\hskip1pt$\sim$}}
    \raise1pt\hbox{$>$}}}
\def\sqr#1#2{{\vcenter{\vbox{\hrule height.#2pt
         \hbox{\vrule width.#2pt height#1pt \kern#1pt
         \vrule width.#2pt}
         \hrule height.#2pt}}}}

\newcommand{\beq}{\begin{equation}}
\newcommand{\eeq}{\end{equation}}
\newcommand{\bea}{\begin{eqnarray}}
\newcommand{\eea}{\end{eqnarray}}

\begin{document}

    \title{Polarization of the CMB in the Standard Model Extension}
\author{Iman Motie}
\affiliation{Université de Toulouse, UPS-OMP, IRAP, F-31400 Toulouse, France }	
\author{ Jafar Khodagholizadeh}
\affiliation{Department of Physics Education, Farhangian University, P.O. Box 14665-889, Tehran, Iran}	
\author{ S. Mahmoudi}
\affiliation{School of Particles and Accelerators, Institute for Research in
 Fundamental Sciences (IPM), P. O. Box 19395-5531, Tehran, Iran}	
\author{Brahim Lamine}
\affiliation{Université de Toulouse, UPS-OMP, IRAP, F-31400 Toulouse, France }	
\author{Alain Blanchard}
\affiliation{Université de Toulouse, UPS-OMP, IRAP, F-31400 Toulouse, France }	

        
\begin{abstract}
   
In standard cosmology, Cosmic Microwave Background photons near the last scattering surface exhibit only linear polarization due to Compton scattering, leading to the assumption that primordial circular polarization is negligible. However, the physics of Lorentz violation (LV), associated with specific operators, can influence these polarization characteristics. This study employs the Boltzmann equation within the framework of the Standard Model Extension (SME) to explore how the background LV tensor $ K_{AF} $ can induce circular polarization in CMB radiation. By computing the transformation of linear polarization into circular polarization and utilizing the Faraday conversion angle, we derive a bound for $ K_{AF} $ on the order of $ 10^{-41} \, \text{GeV} $, aligning with recent findings. Additionally, we consider the total pure photon terms within the SME, demonstrating that LV in the presence of scalar perturbations can also generate cosmic birefringence (CB) in the CMB radiation. Through analysis of best-fitting CB angles, we establish a more stringent bound of approximately $ 10^{-32} \, \text{GeV} $ for $ K_{F} $.
 
\end{abstract}


\maketitle

\section{Introduction}
\label{sec1}
The cosmic microwave background (CMB) radiation, a pervasive glow of microwave radiation, fills the universe and stands as crucial evidence supporting the Big Bang theory. Its detection laid the foundation for the standard model of cosmology, $\Lambda$CDM, which explains the universe’s evolution and structure. During the tight-coupling regime preceding the last scattering of CMB radiation, rapid Compton scattering, relative to the cosmological expansion timescale, kept the cosmic radiation field unpolarized. As the universe cooled, a small degree of linear polarization arose during recombination due to the interaction between free-streaming photons and free electrons via Compton scattering, while circular polarization was not produced in this process \cite{KOSOWSKY199649, Planck2018, hu1997cmb}.

To explore physics beyond this standard scenario, investigating the possibility of circular polarization in CMB radiation becomes essential. The detection of circular polarization would imply new physics, potentially revealing processes from the early universe, particularly from the epoch of cosmic inflation. Many inflationary models predict distinct patterns of circular polarization in the CMB radiation, making its detection valuable for validating or refining these models and for uncovering insights into the universe's earliest moments \cite{Baumann_2009, ALEXANDER2019197, Kamionkowski1999, sarkar2022detecting}. Furthermore, the study of CMB polarization offers a promising probe for early-universe magnetogenesis and the nature of primordial magnetic fields, which may hold clues to symmetry-breaking processes that occurred shortly after the Big Bang \cite{Zucca:2016iur,Khodagholizadeh:2014nfa,Khodagholizadeh:2019het}.

From a theoretical standpoint, various secondary processes could induce circular polarization in the CMB radiation at low levels. Linear polarization in the CMB, influenced by background fields, particle scatterings, and temperature fluctuations, can undergo rotation and convert into circular polarization—a phenomenon described by Faraday rotation (FR) and Faraday conversion (FC), captured by the evolution of the Stokes parameter $ V $ \cite{Jones},\cite{Ko},
\begin{eqnarray}
	\dot V = 2U \frac{d}{dt}(\Delta \Phi_{FC}), \label{fc}
\end{eqnarray}
where $ \Delta \Phi_{FC} $ represents the FC phase shift \cite{COORAY20031}. This process can convert linear polarization into circular polarization within structures such as galaxy clusters \cite{COORAY20031} and in the relativistic plasma remnants of Population III stars \cite{Yaghoobpourb2019}, \cite{tashiro}. In addition, Faraday conversion plays an essential role in understanding magnetized cosmic structures, as it enables the study of magnetic field distributions across galactic and extragalactic environments.

In this context, observations from LOFAR (Low Frequency Array) are pivotal for investigating cosmic magnetic fields, although their capabilities are shaped by frequency range and resolution. Operating between 10–240 MHz, LOFAR effectively measures Faraday depths up to $±100 rad/m$, although limited frequency resolution constrains its ability to detect very high Faraday depths. For studies of galactic and extragalactic foregrounds, LOFAR achieves exceptional precision in low Rotation Measure (RM) observations, reaching accuracies as fine as ±0.1 rad/m² for nearby or low-RM sources. However, its sensitivity typically plateaus around 200–300 rad/m² for more distant or complex regions. Despite these limitations, LOFAR excels in revealing detailed magnetic structures in regions with low Faraday depths, offering valuable insights into cosmic magnetism \cite{OSullivan2023, vanEck2023}.

Beyond Faraday effects, several alternative mechanisms could generate circular polarization in the CMB. These include photon-photon interactions in neutral hydrogen \cite{Sawyer}, primordial magnetic fields \cite{Giovannini, khodam}, Lorentz-invariance-violating operators \cite{Kosteleck, Alexander, khodam}, scattering from the cosmic neutrino background \cite{Mohammadi}, axion-like pseudoscalar particles \cite{Finelli,Mahmoudi:2024wjy}, the impact of vector dark matter \cite{ModaresVamegh2019}, interactions with sterile neutrino dark matter \cite{Haghighat2020}, and nonlinear photon interactions via effective Euler–Heisenberg Lagrangians \cite{Motie2, motie, mohammadimotie1, mohammadimotie2}. The presence of these exotic mechanisms could imprint unique signatures on CMB radiation, offering observable consequences of parity violation, quantum gravity effects, or CPT symmetry breaking—all of which remain open questions in fundamental physics. For instance, polarized Compton scattering in large-scale external magnetic fields could generate circular polarization proportional to the temperature anisotropy power spectrum, $ C_{\ell}^{I(S)} $ \cite{Vahedi_2019, Bunn1996distortion}.

Nevertheless, despite this rich theoretical landscape, the predicted levels of circular polarization remain extremely small and difficult to detect with current observational technology. Present experimental bounds constrain circular polarization to $ \frac{\Delta V}{T_{CMB}} < 10^{-4} $, where $ T_{CMB} $ is the CMB temperature \cite{spider, Mainini2013cmb}. This stringent limit complicates the task of distinguishing among different mechanisms \cite{Ade2019}. {\color{black}As a result,} continued advancements in observational sensitivity are essential for probing these low polarization levels and refining constraints on early-universe physics.

Among the most compelling theoretical approaches, the idea of Lorentz symmetry violation presents a pathway for generating circular polarization. Lorentz symmetry is a foundational aspect of relativity, yet in some theoretical frameworks, it may be systematically broken. The Standard Model Extension (SME) incorporates Lorentz- and CPT-violating terms and predicts observable effects such as cosmic birefringence and circular polarization in the CMB \cite{Caloni_2023, lehnert2015vacuum, bernardis2022cmb}. Studies within the SME framework offer theoretical predictions that guide observational searches for subtle LV signatures, including shifts in polarization angles and asymmetries in CMB spectra.

Furthermore, astrophysical polarization studies offer complementary avenues for testing Lorentz-violating effects. For example, Ref.~\cite{Wei_2019} examines polarization evolution in gamma-ray bursts (GRBs) under Lorentz invariance violation, while multiwavelength polarization measurements of blazars provide constraints on vacuum birefringence through frequency-dependent polarization angle shifts \cite{zhou2021constraints, kislat2019searches}. Lorentz violation could also affect the amplitudes of left- and right-handed gravitational wave modes, resulting in nonzero circular polarization linked to parity-violating parameters.

Motivated by these considerations, this article focuses on the generation of circular polarization through Lorentz-violating interactions within the SME framework. By studying photon interactions with an external Lorentz-violating field, we aim to gain insights into physics beyond the Standard Model. 

This paper is organized as follows: in Sec.~\ref{sec2}, we review the Stokes parameters and Boltzmann equation governing time evolution, and introduce the photon sector of the SME Lagrangian. In Sec.~\ref{seclv}, we compute the evolution of Stokes parameters due to the $ K_{AF} $ interaction term and analyze its contribution to circular polarization and Faraday conversion. Section~\ref{sec4} extends the analysis to include the combined effects of $ K_{AF} $ and $ K_F $. Finally, in Sec.~\ref{conclusion}, we examine the generation of cosmic birefringence due to the $ K_F $ term and conclude with a summary of our findings.


\section{Stokes parameters and Boltzmann equation}
\label{sec2}

To study the polarization state of electromagnetic radiation, including the CMB radiation, it is necessary to use the Stokes parameters. In a general mixed state of photons, the density matrix ($\rho$) in the polarization state space encodes the intensity and polarization of the photon ensemble as follows:
\begin{eqnarray}
 \rho\!\!\!\!\!\!&&=\frac{tr(\rho)}{2}\left(
                       \begin{array}{cccc}
                         I+Q & U-i V \\
                          U+iV & I-Q  \\
                        \end{array}
                     \right)\label{matrix}   =\frac{tr(\rho)}{2}(\mathbf{I}{1} +Q\sigma_3+U\sigma_1+V\sigma_2)
\end{eqnarray}

Here, $\sigma_i$ are the Pauli spin matrices, and $ I $ denotes the total photon intensity. The parameters $ Q $ and $ U $ represent the linear polarization intensities, while $ V $ indicates the difference between left- and right-handed circular polarization intensities. For unpolarized radiation, we have $ Q = U = V = 0 $. 
To track the dynamics of polarization, the time evolution of these Stokes parameters is governed by the Boltzmann equation, and we follow the approach outlined in Refs.~\cite{KOSOWSKY199649,Alexander}.

A free photon gauge field $ A_\mu $, expressed in terms of plane wave solutions and formulated in the Coulomb gauge, can be written as \cite{peskin}:
\begin{eqnarray}
  A^\mu(x) =\int \frac{d^3k}{(2\pi)^3 2k_0}
\times 
[a_i(k)\epsilon_{i}^\mu(k) e^{-ik·x} +  a^\dagger_i (k)
\epsilon^{*\mu}_{i}(k)
e^{ik·x}],\label{photon}  
\end{eqnarray}
where $ \epsilon_i^\mu(k) $ are the polarization four-vectors. The index $ i = 1, 2 $ labels the two transverse polarizations of a free photon with four-momentum $ k $, and $ k_0 = |\boldsymbol{k}| $. Note that the polarization vectors $ \epsilon_{s\mu} $ are assumed to be real here. The creation ($ a_i^\dagger(k) $) and annihilation ($ a_i(k) $) operators satisfy the canonical commutation relation:
\begin{eqnarray}
    [a_i(k),a^\dagger_j(k')] = (2\pi)^3 2k_0\,\delta_{ij}\,\delta^3(\boldsymbol{k}-\boldsymbol{k}').
    \label{anti}
\end{eqnarray}

With this quantization setup, the density matrix elements $ \rho_{ij} $ are related to the number operator $ \mathcal{\hat D}_{ij}({\bf k}) = \hat a_i^\dagger({\bf k})\hat a_j ({\bf k}) $ as follows:
\begin{eqnarray}
\label{eq:3.17}
\langle \mathcal{\hat D}_{ij}({\bf k}) \rangle = (2\pi)^3\, 2k^0\, \delta^{(3)}(0)\, \rho_{ij}({\bf k}).  
\end{eqnarray}

In addition, the time evolution of the number operator is determined by the generalized Boltzmann equation:
\begin{equation}
\label{gb0}
        (2\pi)^3 \delta^3(0)\, 2k^0 \frac{d} {dt} \rho_{ij}(0,{\bf k})=i\langle[\mathcal{\hat H}_{\text{int}}(0),
        \mathcal{\hat D}_{ij}({\bf k})]\rangle -  \frac{1}{2}\int_{-\infty}^\infty dt'
        \langle[  \mathcal{\hat H}_{\text{int}}(t'),[\mathcal{\hat H}_{\text{int}}(0),\mathcal{\hat D}_{ij}({\bf k})]]\rangle,
\end{equation}
where the interaction Hamiltonian is defined as $ \mathcal{\hat H}_{\text{int}}(0) = -\delta \mathcal{L} $. The first term on the right-hand side represents the forward scattering contribution, while the second term corresponds to a higher-order collision term, which becomes significant if the first term vanishes.
At this stage, we focus on the possible effects of Lorentz-violating (LV) interactions on the CMB polarization.

The concept of small violations of Lorentz and CPT invariance originates from more fundamental theories. In particular, the idea of Lorentz symmetry violation emerged in string theory \cite{lvs, lvs2}, where the non-local nature of strings can alter the Lorentz properties of the vacuum. Although the relevant energy scale may appear discouragingly high, spontaneous symmetry breaking provides a mechanism whereby such violations can manifest in low-energy effective theories that reduce to the Standard Model (SM). Notably, high-precision experiments could detect even minute residual effects from such violations.

In recent years, various theoretical frameworks have been proposed to study Lorentz violation at the SM scale. However, many of these approaches either demand fundamental revisions of quantum field theory or remain difficult to investigate. A particularly compelling framework is the Standard Model Extension (SME), which preserves experimental consistency and incorporates spontaneous symmetry breaking in a quantizable way \cite{Kosteleck,Kosteleck2}. In this model, the vacuum acquires non-zero expectation values described by Lorentz tensors. If these expectation values are constant (i.e., spacetime-independent), translational symmetry remains intact, ensuring the conservation of energy and momentum.

It is important to emphasize that even with spontaneous symmetry breaking, Lorentz symmetry continues to be a foundational aspect of the theory. Specifically, the SME remains invariant under observer (passive) Lorentz transformations. This implies that any non-zero vacuum expectation value leads only to particle (active) Lorentz violation, not observer Lorentz violation. In practical terms, this means that the physical laws remain unchanged under coordinate transformations, but local rotations or boosts of particles or fields can lead to measurable effects.

With this theoretical background, the SME Lagrangian incorporates all the usual SM terms along with the most general set of Lorentz-violating (LV) operators that satisfy gauge invariance under $ SU(3) \times SU(2) \times U(1) $ and retain power-counting renormalizability:
\begin{eqnarray}
\mathcal{L}_{SME} = \mathcal{L}_{SM} + \delta \mathcal{L},
\end{eqnarray}
where $ \delta\mathcal{L} $ includes LV terms. These terms are constructed so that part of each term plays the role of a coupling coefficient—often a constant tensor field with spacetime indices—while the remaining part involves SM fields and possibly their derivatives. In the fermionic sector, these interactions can also include gamma matrices.
For the purposes of this study, we restrict our attention to the photon sector of the quantum electrodynamics (QED) part of the SME. The corresponding Lagrangian is given by \cite{Kosteleck,Kosteleck2, motiecar}:
\begin{eqnarray}
\label{QEDSME}
\mathcal{L}^{photon}_{\text{QED}} = -\frac{1}{4}F^{\mu \nu}F_{\mu \nu} + \frac{1}{2}(k_{AF})^\alpha \epsilon_{\alpha \beta\mu\nu}A^{\beta}F^{\mu\nu} - \frac{1}{4}(k_F)_{\alpha \beta\mu\nu}F^{\alpha\beta}F^{\mu\nu},
\end{eqnarray}
where $ F_{\mu \nu} = \partial_\mu A_\nu - \partial_\nu A_\mu $ is the electromagnetic field strength tensor. This Lagrangian consists of the standard Maxwell term, a CPT-odd Lorentz-violating term with coefficient $ k_{AF} $ (dimension of mass), and a CPT-even term with dimensionless coefficient $ k_F $.

\section{Generation and Evolution of Circular Polarization by $K_{AF}$} 
\label{seclv}

In the present section, we examine the evolution of the Stokes parameters using the Boltzmann equation (\ref{gb0}), based on the Lagrangian provided in Eq.~(\ref{QEDSME}). Notably, we have previously addressed the final term involving $k_F$ in Ref.~\cite{khodam}. In contrast, our current approach focuses on the CPT-odd term with coefficient $k_{AF}$. We calculate the evolution of the Stokes parameters and assess its contribution to the generation of circular polarization in the CMB radiation. Subsequently, we will consider the combined influence of both Lorentz-violating terms in Eq.~(\ref{QEDSME}) to analyze their collective impact on cosmic birefringence in Section~\ref{seclv}.

The interaction Hamiltonian is given by
\begin{eqnarray}
    \label{eq:11}
        \mathcal{\hat H}_{\text{int}}(t)&=&-\frac{1}{2} \int d^3 {\bf x} \,
         \:(k_{AF})^\alpha\epsilon_{\alpha \beta\mu\nu}A^{\beta}F^{\mu\nu}\nonumber\\
        &=&-\frac{1}{2} \int d^3 {\bf x} \,
         \:(k_{AF})^\alpha\epsilon_{\alpha \beta\mu\nu}A^{\beta}(\partial_\mu A_\nu - \partial_\nu A_\mu).
\end{eqnarray}
By substituting the definition of the free photon gauge field from Eq.~(\ref{photon}) and applying the canonical commutation relations in Eq.~(\ref{anti}), together with the expectation values given in Eq.~(\ref{eq:3.17}), we can derive the time evolution of the density matrix as:
\begin{eqnarray}
    \label{eq:12}
        \frac{d} {dt} \rho_{ij}({\bf k})=
      \chi \epsilon_{l}^\beta \epsilon^{*\nu}_{s}
      \left[\delta_{sj}\rho_{il}({\bf k}) - \delta_{li}\rho_{sj}({\bf k}) - \delta_{si}\rho_{lj}({\bf k}) + \delta_{lj}\rho_{is}({\bf k})\right],
\end{eqnarray}
where $ \chi = \frac{1}{2k^{0}}(k_{AF})^{\alpha}\epsilon_{\alpha \beta\mu\nu}k^{\mu} $. 
Accordingly, the individual components of the density matrix evolve as follows:
\begin{eqnarray}
    \frac{d} {dt} \rho_{11}({\bf k}) &=& 
    \chi(\rho_{12} - \rho_{21})(\epsilon_2^\beta\epsilon_1^\nu + \epsilon_1^\beta\epsilon_2^\nu) = 0,
\end{eqnarray}
\begin{eqnarray}
    \frac{d} {dt} \rho_{22}({\bf k}) &=& 
    -\chi(\rho_{12} - \rho_{21})(\epsilon_2^\beta\epsilon_1^\nu + \epsilon_1^\beta\epsilon_2^\nu) = 0,
\end{eqnarray}
\begin{eqnarray}
    \frac{d} {dt} \rho_{12}({\bf k}) &=& 
    \chi\left[2\rho_{12}(\epsilon_2^\beta\epsilon_2^\nu - \epsilon_1^\beta\epsilon_1^\nu) + (\rho_{11} - \rho_{22})(\epsilon_2^\beta\epsilon_1^\nu + \epsilon_1^\beta\epsilon_2^\nu)\right]\nonumber\\
    &=& 2\chi\rho_{12}(\epsilon_2^\beta\epsilon_2^\nu - \epsilon_1^\beta\epsilon_1^\nu),
\end{eqnarray}
\begin{eqnarray}
    \frac{d} {dt} \rho_{21}({\bf k}) &=& 
    \chi\left[2\rho_{21}(\epsilon_1^\beta\epsilon_1^\nu - \epsilon_2^\beta\epsilon_2^\nu) + (\rho_{22} - \rho_{11})(\epsilon_2^\beta\epsilon_1^\nu + \epsilon_1^\beta\epsilon_2^\nu)\right]\nonumber\\
    &=& 2\chi\rho_{21}(\epsilon_1^\beta\epsilon_1^\nu - \epsilon_2^\beta\epsilon_2^\nu),
\end{eqnarray}
where we have used the antisymmetric property of $\chi$. For further computational details, please refer to the Appendix.
By applying the definition of the Stokes parameters from Eq.~(\ref{matrix}), we obtain their time evolution:
\begin{eqnarray}
    \dot{I} &=& 0, \label{i}
\end{eqnarray}
\begin{eqnarray}
    \dot{Q} &=& 0, \label{q}
\end{eqnarray}
\begin{eqnarray}
    \dot{U}(k) &=& -\frac{i}{k^0}(k_{AF})^{\alpha}\epsilon_{\alpha \beta\mu\nu}k^\mu(\epsilon_2^\beta\epsilon_2^\nu - \epsilon_1^\beta\epsilon_1^\nu)V, \label{u}
\end{eqnarray}
\begin{eqnarray}
    \dot{V}(k) &=& \frac{i}{k^0}(k_{AF})^{\alpha}\epsilon_{\alpha \beta\mu\nu}k^\mu(\epsilon_2^\beta\epsilon_2^\nu - \epsilon_1^\beta\epsilon_1^\nu)U. \label{v}
\end{eqnarray}

As shown in Eq.~(\ref{v}), the derivative $\dot{V}$ is nonzero at first order in the LV parameter $k_{AF}$, which indicates the generation of circular polarization in the CMB radiation.

In particular, the relationship between the Stokes parameters $U$ and $V$ plays a central role. A non-zero $V(k)$ mode can arise from a non-zero $U(k)$, and vice versa. As evident from Eq.~(\ref{v}), $\dot{V}$ is linearly proportional to $U$, implying that a linearly polarized photon ensemble can acquire circular polarization in the presence of the Lorentz-violating $K_{AF}$ term. 
his conversion process can also be understood through the Faraday conversion phase shift $\Delta\Phi_{FC}$, introduced in Eq.~(\ref{fc}). Using Eqs.~(\ref{fc}) and (\ref{v}), one can express this phase shift as:
\begin{eqnarray}
\Delta\phi_{FC} &=& \int dt\, \frac{i}{2k^0}(k_{AF})^{\alpha}\epsilon_{\alpha \beta\mu\nu}k^\mu(\epsilon_2^\beta\epsilon_2^\nu - \epsilon_1^\beta\epsilon_1^\nu). \label{result2}
\end{eqnarray}

To estimate $\Delta\phi_{FC}$, we integrate over the comoving time: 
$\int dt = \int dz /[(1+z)H(z)]$, with redshift $z \in [0, 1000]$. The Hubble parameter is given by 
$H(z) = H_0[\Omega_M(1+z)^3 + \Omega_\Lambda]^{1/2}$, where $\Omega_M \simeq 0.3$, $\Omega_\Lambda \simeq 0.7$, and $H_0 = 72.6$ km/s/Mpc, based on recent Webb telescope data \cite{scolnic2025hubble}. The CMB temperature evolves as $T = T_0(1 + z)$, with $T_0 \approx 2.725\, \text{K}$ \cite{Gawiser2000, Fixsen1996}. We find:
\begin{eqnarray}
\Delta\phi_{FC} &=& \frac{1}{2}(k_{AF})^{\alpha}\epsilon_{\alpha \beta\mu\nu}(\epsilon_2^\beta\epsilon_2^\nu - \epsilon_1^\beta\epsilon_1^\nu) \int_0^{1000} \frac{dz}{H_0(1+z)[\Omega_M(1+z)^3 + \Omega_\Lambda]^{1/2}} \nonumber\\
&\approx& 3 \times 10^{41}\text{GeV}^{-1} \cdot k_{AF}. \label{result2}
\end{eqnarray}

Assuming a phase shift of order unity, the LV parameter $k_{AF}$ can be estimated as:
\begin{eqnarray}
k_{AF} \simeq 10^{-41}\text{GeV}.
\end{eqnarray}

{\color{black}Interestingly,} this value is consistent with earlier observations of astrophysical birefringence, which estimated $2|k_{AF}| \approx 10^{-41}~\text{GeV}$  \cite{Nodland1997}. However, later analyses found no statistically significant signal \cite{Carroll_1997}. 
In the framework of the Standard-Model Extension (SME), signatures of Lorentz violation in electrodynamics can affect CMB anisotropies through both CPT-odd and CPT-even renormalizable operators. The limit on the CPT-odd coefficient is 
$|k_{AF}| < 7.4 \times 10^{-45}\approx\text{GeV}$ at $95\%$ confidence level, which is three orders of magnitude more stringent than our result~\cite{Caloni_2023}.
By classifying all gauge-invariant Lorentz- and CPT-violating terms in the quadratic Lagrangian density for the effective photon propagator, and comparing with BOOMERANG (B03) data~\cite{Montroy2006}, the CMB polarization sensitivity to Lorentz-violating operators is shown to depend on the mass dimension  $d$ . For the case  $d = 3 $, an upper bound of approximately  $(15 \pm 6) \times 10^{-43}~\mathrm{GeV}$ has been established~\cite{Kostelecky2009, Kostelecky2007}.
Additionally, deviations from a zero photon mass in the de Broglie-Proca (dBP) theory or Lorentz symmetry violation (LSV) in the SME framework---tested using solar wind observations imply an upper bound of 
$|\mathbf{k}_{AF}| < 1.03 \times 10^{-26}~\mathrm{GeV}$ \cite{Spallicci2024}. In this experimental approach, the photon mass is first inferred, and then the LSV parameter is constrained indirectly.
 
\section{The Cosmic Birefringence due to LV Interaction}
\label{sec4}

This section aims to investigate the role of Lorentz-violating (LV) interactions in the cosmic birefringence (CB) effect, which refers to the rotation of the linear polarization plane of CMB photons as they propagate through spacetime. Specifically, when the polarization plane is rotated uniformly by an angle $\beta$ over the entire sky, the observed polarization components transform as $Q^{\mathrm{o}} \pm iU^{\mathrm{o}} = (Q \pm iU)e^{\pm 2i\beta}$, where the superscript “o” denotes the observed value, and $Q \pm iU$ on the right-hand side represent the intrinsic values at the last scattering surface. Here, $\beta$ is known as the birefringence angle, and its most recent reported value is $\beta = 0.30^\circ \pm 0.11^\circ \simeq 5 \times 10^{-3}$ \cite{Minami:2019ruj, PhysRevLett.128.091302}.

{\color{black}As discussed earlier,} the pure photon sector of the SME Lagrangian given in Eq.~(\ref{QEDSME}) includes two LV terms, with coefficients $k_F$ and $k_{AF}$. In Eqs.~(\ref{i})–(\ref{v}), we have demonstrated how the Stokes parameters for CMB radiation evolve over time due to the influence of the $k_{AF}$ term. {\color{black}Additionally,} in our previous work \cite{khodam}, we calculated the effects of the $k_F$ term. 
{\color{black}Now, combining both contributions,} we analyze the complete LV effect (within the photon sector) on cosmic birefringence. The full time evolution of the Stokes parameters is given by:
\begin{eqnarray}
     \dot{I} &=& 0, \label{i2}
\end{eqnarray}
\begin{eqnarray}
   \dot{Q} &=& -\frac{16}{k^{0}} (k_{F})_{\mu\nu\alpha\beta}k^{\alpha}k^{\mu}(\epsilon _{1}^\beta \epsilon^{*\nu}_{2}-\epsilon _2^\beta \epsilon^{*\nu}_{1}) V, \label{q2}
\end{eqnarray}
\begin{eqnarray}
   \dot{U}(k) &=& -\frac{i}{k^{0}} (k_{AF})^{\alpha}\epsilon_{\alpha \beta\mu\nu}k^{\mu}(\epsilon_2^\beta\epsilon_2^\nu - \epsilon_1^\beta\epsilon_1^\nu)V \nonumber\\
   && -\frac{4}{k^{0}}(k_{F})_{\mu\nu\alpha\beta}k^{\alpha}k^{\mu}(\epsilon _{2}^\beta \epsilon^{*\nu}_{1}-\epsilon _{1}^\beta \epsilon^{*\nu}_{2}) Q, \label{u2}
\end{eqnarray}
\begin{eqnarray}
   \dot{V}(k) &=& \frac{i}{k^{0}}(k_{AF})^{\alpha}\epsilon_{\alpha \beta\mu\nu}k^{\mu}(\epsilon_2^\beta\epsilon_2^\nu - \epsilon_1^\beta\epsilon_1^\nu)U \nonumber\\
   && -\frac{4i}{k^{0}}(k_{F})_{\mu\nu\alpha\beta}k^{\alpha}k^{\mu} \Big[ (\epsilon _{1}^\beta \epsilon^{*\nu}_{1}-\epsilon _{2}^\beta \epsilon^{*\nu}_{2}) U - (\epsilon _{1}^\beta \epsilon^{*\nu}_{2} + \epsilon _{2}^\beta \epsilon^{*\nu}_{1}) Q \Big]. \label{v2}
\end{eqnarray}

{\color{black}According to} scalar-mode metric perturbations \cite{BondEfstathiou1984}, and using the calculations in the previous section, Eqs.~(\ref{q2}) and (\ref{u2}) imply that LV interactions modify the time evolution of the linear polarization of CMB radiation as follows:
\begin{equation}\label{BE3}
\dfrac{d}{d\eta}\Delta^{\pm \text{S}}_{\text{P}} + iK\mu\Delta^{\pm \text{S}}_{\text{P}} = \dot{\tau_{\text{e}}} \left[ -\Delta^{\pm \text{S}}_{\text{P}} - \dfrac{1}{2}(1 - P_2(\mu)) \Pi \right] \mp i a(\eta)\dot{\tau}_{\text{\tiny{LIV}}}^{Q}\Delta^{\text{S}}_{\text{Q}} + a(\eta)\dot{\tau}_{\text{\tiny{LIV}}}^{\pm V}\Delta^{\text{S}}_{\text{V}},
\end{equation}
where $\Delta^{\pm}_{\text{P}} = Q \pm iU$, and $ \Pi \equiv \Delta^S _{I2} + \Delta^S _{P2} + \Delta^S _{P0} $. The LIV contributions are expressed as:
\begin{equation}
\dot{\tau}_{\text{\tiny{LIV}}}^{Q} = \frac{4}{k^{0}}(k_{F})_{\mu\nu\alpha\beta}k^{\alpha}k^{\mu}(\epsilon _{2}^\beta \epsilon^{*\nu}_{1} - \epsilon _{1}^\beta \epsilon^{*\nu}_{2}),
\end{equation}
\begin{equation}
\dot{\tau}_{\text{\tiny{LIV}}}^{+ V} = -\frac{16}{k^{0}}(k_{F})_{\mu\nu\alpha\beta}k^{\alpha}k^{\mu}(\epsilon _{1}^\beta \epsilon^{*\nu}_{2} - \epsilon _2^\beta \epsilon^{*\nu}_{1}) + \frac{1}{k^{0}}(k_{AF})^{\alpha}\epsilon_{\alpha \beta\mu\nu}k^{\mu}(\epsilon_2^\beta\epsilon_2^\nu - \epsilon_1^\beta\epsilon_1^\nu),
\end{equation}
\begin{equation}
\dot{\tau}_{\text{\tiny{LIV}}}^{- V} = -\frac{16}{k^{0}}(k_{F})_{\mu\nu\alpha\beta}k^{\alpha}k^{\mu}(\epsilon _{1}^\beta \epsilon^{*\nu}_{2} - \epsilon _2^\beta \epsilon^{*\nu}_{1}) - \frac{1}{k^{0}}(k_{AF})^{\alpha}\epsilon_{\alpha \beta\mu\nu}k^{\mu}(\epsilon_2^\beta\epsilon_2^\nu - \epsilon_1^\beta\epsilon_1^\nu).
\end{equation}

{\color{black}Focusing on} the dominant contribution in Eq.~(\ref{BE3}), the equation simplifies to:
\begin{equation}\label{BE31}
\dfrac{d}{d\eta}\Delta^{\pm \text{S}}_{\text{P}} + iK\mu\Delta^{\pm \text{S}}_{\text{P}} \approx \dot{\tau_{\text{e}}} \left[ -\Delta^{\pm \text{S}}_{\text{P}} - \dfrac{1}{2}(1 - P_2(\mu)) \Pi \right] \mp i a(\eta)\dot{\tau}_{\text{\tiny{LIV}}}^{Q} \Delta^{\text{S}}_{\text{Q}},
\end{equation}
where $\dot{\tau}_{\text{\tiny{LIV}}}^{Q}$ quantifies the strength of the LV contribution responsible for inducing the CB effect. 

The CB angle is approximately related to the effective optical depth by \cite{Khodagholizadeh:2023aft}
\begin{equation}
\beta \approx \frac{1}{2} \tau_{\text{LIV}}^Q ,
\label{beta}
\end{equation}
where
\begin{eqnarray}
\tau_{\text{LIV}}^Q(z) &=& \int_0^z \frac{dz'}{(1 + z') H(z')} \dot{\tau}_{\text{LIV}}^Q \nonumber\\
&=& \int_0^z \frac{dz'}{(1 + z') H_0 \left[ \Omega_M(1 + z')^3 + \Omega_\Lambda \right]^{1/2}} \dot{\tau}_{\text{LIV}}^Q. \label{BE9}
\end{eqnarray}

{\color{black}By taking} $ k \simeq 10^{-4} $ and $ H_0^{-1} = 6 \times 10^{46}\,\text{GeV}^{-1} $, we find
\begin{eqnarray}
\tau_{\text{LIV}}^Q(z) &\approx& 7.2 \times 10^{29} K_F,
\end{eqnarray}
and thus,
\begin{equation}
\beta \approx 3.6 \times 10^{29} K_F.
\end{equation}

This yields a bound of $ K_F \simeq 10^{-32} $. {\color{black}Notably,} this result is one order of magnitude stronger than previous constraints based on Lorentz violation signatures in CMB anisotropies, which found that the CPT-even coefficient must satisfy $ K_F < 2.3 \times 10^{-31} $ at $95\%$ confidence level \cite{Caloni_2023}.


\section{Conclusions} \label{conclusion}
In this work, we have examined the evolution of the Stokes parameters resulting from the interaction of photons with a background Lorentz violation field. It is established that nonzero backgrounds can produce circular polarization in the CMB radiation. Within the pure sector of the SME, we identified two distinct terms: one with CPT-odd $k_{AF}$ coefficients and the other with CPT-even $K_F$ coefficients. Currently, there is no confirmed experimental evidence supporting Lorentz violation, which includes existing experimental measurements and theory-derived limits on these SME coefficients.

For the case of the $k_{AF}$ term, we calculated the time derivative of the Stokes parameter using the Boltzmann equation. Our results indicate that a nonzero $\dot{V}$ guarantees the generation of circular polarization via the Faraday conversion angle. We found that the time evolution of the Stokes parameter $V$ is linearly proportional to the dimensionless Lorentz parameter $k_{AF}$, which is comparable to results obtained in \cite{khodam} where the CPT-even $K_F$ term of the Lagrangian was considered as the interaction term. We also estimated the Faraday phase conversion in this LV case, yielding an upper bound on the Lorentz parameter $k_{AF} \approx 10^{-41}, \text{GeV}$. This result aligns with current bounds established within the minimal SME framework and is consistent with systematic observations of the rotation of the plane of polarization of electromagnetic radiation.
Despite these findings, current instrumental techniques struggle to detect circular polarization in the CMB due to sensitivity limitations. The experimental bound on the $V$ modes, as reported in \cite{Partridge1988}, constrains it to $V \leq 1 \, \mu\text{K}$. However, post-Planck CMB polarization experiments are actively being developed \cite{Baumann_2009}. Understanding whether the CMB exhibits circular polarization and what physical insights can be gleaned from its measurement is crucial. If new experimental devices can achieve polarimetric sensitivity beyond the $\mu K$ range and effectively manage systematic errors, the resulting observations could potentially uncover physics at energy scales beyond the capabilities of current particle accelerators, thus providing a window into new physics.

Additionally, we summarized our results regarding the time evolution of the Stokes parameters for $K_{AF}$ and $K_F$, previously computed in \cite{khodam}, to assess their contributions to the cosmic birefringence effect. In these cases, $\dot{U}$ and $\dot{Q}$ are critical, demonstrating that while both parameters could generate birefringence, the influence of $K_F$ is more significant than that of $k_{AF}$. Using experimental data on cosmic birefringence angles, we established an upper bound on the dimensionless Lorentz parameter $k_F \approx 10^{-32} \, \text{GeV}$, which not only aligns with but is also stronger than existing bounds on this parameter.
Looking ahead, future research in this area could focus on several key avenues. First, the development of advanced observational techniques will be critical for detecting circular polarization in the CMB. The design of new instruments with enhanced sensitivity could enable direct measurements of circular polarization, thus testing our theoretical predictions. Notable upcoming projects include the CMB-S4 experiment, which aims to significantly enhance sensitivity and resolution in CMB observations. Similarly, the Simons Observatory \cite{SimonsObservatory2025}and PICO \cite{PICO2019}(Polarized Intensity and Cosmic Origins) experiment are also anticipated to provide valuable data on CMB polarization, potentially revealing circular polarization signatures.

Furthermore, future analyses could incorporate upcoming data from next-generation CMB experiments, which may provide more stringent constraints on LV parameters. The Euclid Space Telescope, scheduled for launch, will also complement these observations by studying the universe's large-scale structure and dark energy effects, indirectly informing models of Lorentz violation.

Additionally, further theoretical studies are needed to explore the implications of LV on other cosmological phenomena, such as gravitational wave propagation and the behavior of dark matter. Investigating the interplay between LV and cosmic inflationary models could yield insights into the fundamental physics governing the early universe.

By addressing these research gaps and leveraging next-generation observational technologies, we can deepen our understanding of the cosmos and potentially uncover new physics that challenges or extends the current standard model of cosmology.

\appendix
\section{Appendix }\label{app}
  This section considers the quantum number operator for a system of photon particles
and derives its evolution equation, including local particle interactions. Taking the expectation value of the operator equation gives the Boltzmann equation for the
system's density matrix, which is a generalization of the usual classical Boltzmann equation for particle occupation numbers (the diagonal elements of the density
matrix).

The density operator describing a system of photons is given by
\begin{eqnarray}
    \hat \rho =\int \frac{d^3{\bf p'}}{(2\pi)^3}\rho_{ij}({\bf p'})a^\dagger_i({\bf p'})a_j({\bf p'}),
\end{eqnarray}
where $\rho_{ij}$ is the is the density matrix and the bold momentum variables represent three-momenta.
The particular operator for which we want the equation of motion is the photon number operator 
\begin{eqnarray}
\mathcal{D}_{ij}({\bf p'})=a^\dagger_i({\bf p'})a_j({\bf p'}).
\end{eqnarray}
The expectation value of $\mathcal{D}$ is proportional to the density matrix. 

The Hamiltonian can be written as 
\begin{eqnarray}
\mathcal{\hat H}_{\text{int}}&=&-\frac{1}{2} \int d^3 {\bf x} \,
         \:(k_{AF})^\alpha\epsilon_{\alpha \beta\mu\nu}A^{\beta}(\partial_\mu A_\nu-\partial_\nu A_\mu),
\end{eqnarray}
where by using (\ref{photon}),
\begin{equation}
    \label{eq:10}
       \partial_{\mu} A_\nu(x) = \int d{\bf k}(-ik_{\mu}) \left[ \hat a_{s}(k) \epsilon _{s\nu}(k)
        e^{-ik\cdot x}-\hat a_{s}^\dagger (k) \epsilon^* _{s\nu}(k)e^{ik\cdot x}
        \right], 
\end{equation}
then 
\begin{eqnarray}       
        \mathcal{\hat H}_{\text{int}} &=&\frac{-i}{2}(k_{AF})^\alpha\epsilon_{\alpha \beta\mu\nu}\int d^3x\int \frac{d^3{\bf p}}{2p^0(2\pi)^3}\int \frac{d^3{\bf p'}}{2p'^0(2\pi)^3}\Sigma_{s,l}
        \nonumber\\
        &\times&
        \bigg\{[a_l({\bf p'})a_s^\dagger({\bf p})e^{i(p-p').x
        }
        -a_l^\dagger({\bf p'})a_s({\bf p})e^{-i(p-p').x}]p^\mu\epsilon_l^\beta(p')\epsilon_s^\nu(p)
        \nonumber\\
        &+&
        [a_l^\dagger({\bf p'})a_s({\bf p})e^{-i(p-p').x}-a_l({\bf p'})a_s^\dagger({\bf p})e^{i(p-p').x}]p^\nu\epsilon_l^\beta(p')\epsilon_s^\mu(p)\bigg\}.
\end{eqnarray}

The  forward
scattering term in Eq.~(\ref{gb0}) is given as
\begin{eqnarray}
[\mathcal{\hat H}_{\text{int}}(0),
        \mathcal{\hat D}_{ij}({\bf k})]&=&\frac{-i}{2}(k_{AF})^\alpha\epsilon_{\alpha \beta\mu\nu}\int d^3x\int \frac{d^3{\bf p}}{2p^0(2\pi)^3}\int \frac{d^3{\bf p'}}{2p'^0(2\pi)^3}\Sigma_{s,l}
        \nonumber\\&\times&\bigg\{[a_l({\bf p'})a_s^\dagger({\bf p}),a^\dagger_i({\bf k})a_j({\bf k})]\{p^\mu \epsilon_l^\beta(p')\epsilon_s^\nu(p)-p^\nu \epsilon_l^\beta(p')\epsilon_s^\mu\}e^{i(p-p').x}\nonumber\\
        &+&
        [a_l^\dagger({\bf p'})a_s({\bf p}),a^\dagger_i({\bf k})a_j({\bf k})]\{-p^\mu \epsilon_l^\beta(p')\epsilon_s^\nu(p)+p^\nu \epsilon_l^\beta(p')\epsilon_s^\mu\}e^{-i(p-p').x}
        \bigg\},
\end{eqnarray}
and the expectation value of the forward scattering term is given by
\begin{eqnarray}
    i\langle[\mathcal{\hat H}_{\text{int}}(0),
        \mathcal{\hat D}_{ij}({\bf k})]\rangle&=& \epsilon_{\alpha \beta\mu\nu}\,\,k_{AF}^\alpha \,\,k^\mu\,(2\pi )^3\, \delta^3 (0)\Sigma_{l,s}\epsilon_l^\beta(k)\epsilon_s^\nu(k)
        \nonumber\\&\times&\bigg\{
        \delta_{sj}\rho_{il}(k)-\delta_{li}\rho_{sj}(k)-\delta_{si}\rho_{lj}(k)+\delta_{lj}\rho_{is}(k)
        \bigg\}
\end{eqnarray}

\end{document}